# Signal Enhancement and Background Suppression Using Interference and Entanglement

## Keith Kastella[*] and Ralph S. Conti[+]

[*]*SRI International, 2100 Commonwealth Blvd.,*
*Suite 210, Ann Arbor, MI 48105, Keith.Kastella@SRI.com*
*[+]General Dynamics Advanced Information Systems, Michigan Research & Development*
*center, 1200 Joe Hall Dr. Ypsilanti, MI 48197 ralph.conti@gd-ais.com and Physics*
*Department, University of Michigan, Ann Arbor, MI 48109, RConti@UMich.edu*

**Abstract:** We describe a two-photon absorption process that is excited by entangled pairs but not non-entangled pairs with the same energy and polarization. Photon states can be selected so that in the non-entangled process, there is destructive interference between different orders of absorption and intermediate state contributions. A non-zero entangled absorption cross section is obtained by varying the entanglement time and/or pair delay parameters. As an example, the destructively interfering energy and polarization states and the resulting entangled absorption cross section for Rb $5S_{1/2} \rightarrow 5D_{3/2}$ transitions are computed. This effect can be used to construct an entangled photon detector with applications in sensing, cryptography, and lithography.

The past two decades have seen an explosion in quantum entanglement research in areas such as quantum computation[1], quantum cryptography[2], and lithography[3]. Entangled photons, entangled atoms and entangled atom-photon systems have all been demonstrated. A number of sensing concepts that employ entangled photons pairs (biphotons) have been demonstrated[4] and have engendered similar concepts that rely on non-quantum correlations[5]. Quantum imaging is beginning to emerge as a field.[6] In most of this work, the detection scheme relies on coincidence measurements between photon pairs. The use of coincidence detection presents challenges in experiments of this nature. The experiment rate is limited by the speed of the electronics. The illumination intensity must be low enough that no more than a single signal/idler pair arrives within the coincidence window. If multiple pairs arrive at the detectors within the same gate time, signal and idler photons derived from different pump photons can result in unwanted coincidence measurements, contributing to noise in the system.

Recently, a number of authors have shown that upconversion processes can act as a detector.[7] This has the advantage of allowing characterization of the photon wave packets on femtosecond time scales. In this Letter we show that this type of detection scheme can also be designed to respond preferentially to entangled pairs, using quantum interference effects to suppress false detections from otherwise identical non-entangled pairs. Thus we can create a "biphoton detecting device (BDD)[8]" that can test for entanglement vs. non-entanglement of photon pairs.

One application of such a device is to target detection. Consider the problem of detecting an object against a strong background. A beam of photons is directed at a possible target location. If no target is present, the sensor detects only background. If a target is present, then the sensor detects both the background and photons reflected from the target. If the background is strong compared with the target, then it can be difficult to differentiate signal photons scattered by the target from background photons. Recently, "quantum illumination"[9] was suggested as a mechanism to differentiate signal photons from background photons with enhanced signal-to-noise achieved due to the phase space correlation between signal and idler photons. The biphoton detection approach provides additional enhancement by using quantum interference for background suppression.

The theory of random two-photon absorption (RTPA) was developed by Goppert-Mayer[10] and extended to a number of multi-photon processes[11]. There is significant structure in the two-photon cross section with strong enhancement when one of the single photon energies approaches an allowed intermediate state transition. Destructive interference between pairs of intermediates was also predicted and observed.[12] The entangled two-photon absorption (ETPA) cross section for atomic systems was derived by [13] and generalized to entangled multi-photon processes by [14], while [15] suggest that entanglement modulation can be used to access virtual state information in atomic systems.

Given the flux $\phi$ of entangled photons incident on a target, both entangled and random absorption processes can occur. The entangled process is linear in $\phi$ while the random process is quadratic. This can be estimated using a probabilistic model from the single-photon cross section $\sigma$, the virtual state lifetime $\tau_v$, the entangled photon correlation time $T_e$ and the correlation area $A_e$. $\tau_v$ is determined by the uncertainty principle from the energy defect $\Delta E$ between the absorbed photon energy and the energy of the closest intermediate atomic state.[16] The biphoton correlation time is an entanglement time derived from the superposition of signal and idler delays.[17,18] If multiple spatial modes of the signal and idler are used, then the correlation area $A_e$ is determined by the momentum entanglement of the biphoton, but if single modes of the signal and idler are captured and transmitted to the target, then the correlation area $A_e$ is the focal spot size of the signal and idler beams. As a result, the total two-photon absorption rate is given by

$$R = R_e + R_r = \sigma_e \phi + \delta_r \phi^2 \qquad (1)$$

where $\sigma_e \sim \sigma^2 \tau_v / 2 A_e T_e$, $\delta_r \sim \sigma^2 \tau_v$ and the biphoton-flux density is $\phi / 2$. Although the entangled linear absorption regime has been demonstrated in targets consisting of up-conversion crystals[7] and organic molecules,[19] it has not yet been demonstrated in atomic vapor targets. In the linear regime, [19] also exhibited non-monotonic modulation with inter-beam delay, indicative of an entangled interference process.

To develop a BDD system, we start with entangled photons produced using spontaneous parametric down-conversion (SPDC). In SPCD, photons from a pump laser beam are split into entangled photon pairs by a non-linear crystal or waveguide, satisfying $\omega_p = \omega_1 + \omega_2$ and $\mathbf{k}_p = \mathbf{k}_1 + \mathbf{k}_2$ where $\omega_{p,1,2}$ and $\mathbf{k}_{p,1,2}$ are the photon frequencies and wave-vectors and $p$, 1 and 2 index the pump, signal, and idler variables respectively.

The electric field distribution for states of definite polarization at the output of the crystal face is given by the biphoton wave function[14]

$$\langle 0|\mathbf{E}_1^{(+)}(\mathbf{r}_1,t_1)\mathbf{E}_2^{(+)}(\mathbf{r}_2,t_2)|II\rangle \equiv \Psi(\mathbf{r}_1,t_1,\mathbf{r}_1,t) \tag{2}$$
$$= v_0 \exp\left(-i\omega_1(t_1 - z_1/c) - i\omega_2(t_2 - z_2/c)\right)$$
$$\times \exp\left(-\left(\frac{x_1 - x_2}{\Delta x}\right)^2 - \left(\frac{y_1 - y_2}{\Delta y}\right)^2\right)\Pi\left(\frac{t_1 - z_1/c - t_2 + z_2/c}{2T_e}\right)$$

where

$$\Pi(t) = \begin{cases} 1, & 0 \le t \le 1 \\ 0, & \text{otherwise.} \end{cases} \tag{3}$$

$\mathbf{E}_i^{(+)}(\mathbf{r}_i,t_i)$ is the interaction picture operator, $T_e = T_1 - T_2 = l(1/u_1 - 1/u_2)/2$, $v_0^2 = \hbar^2\omega_1\omega_2/\left(2\varepsilon_0^2 V_Q c A_e T_e\right)$, $A_e \equiv \pi\Delta x\Delta y/2$, and quantization volume $V_Q = \int d^3r$.

The wave-function (2) is both momentum- and energy-entangled since it is not factorizable in either space or time variables. The non-factorizability is caused by the superposition of the pair creation points within the crystal. Because of dispersion and/or birefringence, different pair-creation points lead to different temporal and spatial separation of the signal and idler rays. The spatial variation scale is large compared with atomic length scales, so the biphoton wave-function can be taken as approximately constant over a single atom. Placing the atom at the coordinate system origin, we set $x_1 = x_2 = 0$ and similarly for $y$ and $z$. The difference between the entangled and non-entangled cross sections is due entirely to the box-car function $\Pi(t)$.

Both RTPA and ETPA use photons with frequencies near $\omega_1^0$ and $\omega_2^0$ to drive the transition from ground state frequency $\varepsilon_g$ to final state $\varepsilon_f$, satisfying $\omega_1^0 + \omega_2^0 = \varepsilon_f - \varepsilon_g$. For a monochromatic pump, the ETPA cross section is[13]

$$\sigma_e = \frac{\pi}{(\hbar c\varepsilon_0)^2 4A_e T_e}\omega_1^0\omega_2^0\delta\left(\varepsilon_f - \varepsilon_i - \omega_1^0 - \omega_2^0\right)|M_e|^2 \tag{4}$$

where

$$M_e = \sum_i \left\{ D_{21}^i \frac{1 - \exp\left[-i\Delta_1^i(T_e + \tau)\right]}{\Delta_1^i - i\kappa_i/2} + D_{12}^j \frac{1 - \exp\left[-i\Delta_2^i(T_e - \tau)\right]}{\Delta_2^i - i\kappa_i/2} \right\} \tag{5}$$

with summation over intermediate states $i$ with width $\kappa_i$, photon indices $j,k = 1,2$, $\Delta_j^i = \varepsilon_i - \varepsilon_g - \omega_j$, and $D_{jk}^i = \langle\psi_f|d_j|\psi_i\rangle\langle\psi_i|d_k|\psi_g\rangle$. The transition dipole matrix element between atomic states $a$ and $b$ with photon polarization $\boldsymbol{\lambda}_j$ is $\langle\psi_b|d_j|\psi_a\rangle$ where the dipole operator is $\mathbf{d} = -e\mathbf{r}$ and $d_j = \boldsymbol{\lambda}_j \cdot \mathbf{d}$. Eq. (5) is a generalization of [13] to include the effect of inter-beam delay $\tau$ and is consistent with [15].

The RTPA cross section is:

$$\delta_r = \frac{\pi}{2(\hbar c \varepsilon_0)^2} \omega_1^0 \omega_2^0 \delta\left(\varepsilon_f - \varepsilon_i - \omega_1^0 - \omega_2^0\right) |M_r|^2 \qquad (6)$$

with matrix element

$$M_r = \sum_i \left\{ \frac{D_{21}^i}{\Delta_1^i - i\kappa_i/2} + \frac{D_{12}^i}{\Delta_2^i - i\kappa_i/2} \right\} \qquad (7)$$

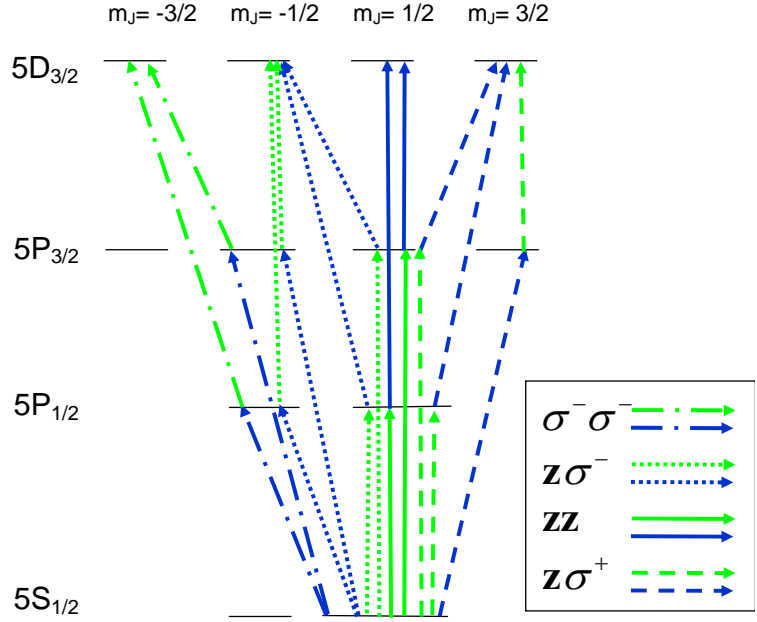

**Figure 1** – Candidate BDD two-photon absorption processes from the $5S_{1/2}$, $m_J = 1/2$ Rb ground state into a single final state (energy levels not to scale).

We now determine how to select photon energies and polarization states so that RTPA destructively interferes in Rb, allowing it to be used as an entanglement-selecting detector. The low-lying fine structure of Rb is shown in Figure 1. The selection rules for dipole transitions are $\Delta L = \pm 1$, and $\Delta M = 0$, $1$ or $-1$ respectively for $\mathbf{z}$-, $\sigma^+$-, and $\sigma^-$ polarized light. For a two-photon transition starting in the $S$, $m_J = +1/2$ ground state, $\Delta L = \pm 1$ requires that the final state be either a D or an S state. The lowest energy options are transitions to the $5D_{3/2}$ states. For transitions that involve more than a single final state, the absorption cross section requires averaging over the final states, making it impossible to solve for destructive interference unless all final states separately interfere. Figure 1 shows 4 candidate transitions from $5S_{1/2}$ into a unique $m_J$ final state with photon polarizations: $\sigma^-\sigma^-$, $\mathbf{z}\sigma^-$, $\mathbf{z}\mathbf{z}$, and $\mathbf{z}\sigma^+$. By symmetry, there are four additional transitions from the $m_J = -1/2$ with $\sigma^+ \leftrightarrow \sigma^-$. The $\mathbf{z}\sigma^-$ and $\mathbf{z}\sigma^+$ transitions require state preparation, such as optical pumping, to eliminate transitions from the $m_J = -1/2$ state. The $\sigma^-\sigma^-$ and $\mathbf{z}\mathbf{z}$ transitions do not require state preparation, since selection rules forbid competing transitions for $\sigma^{-}\sigma^{-}$, while **zz** yields the same result for transitions from either $m_J$ ground state.

To evaluate the ETPA and RTPA transitions, define $\varepsilon_g = 0$ and $\omega_{1,2}^0 = \omega_p / 2 \mp \delta$. There are two fine structure intermediate levels with frequencies $\varepsilon_{P_{1/2}} \equiv \omega_p / 2 - a$ and $\varepsilon_{P_{3/2}} \equiv \omega_p / 2 - b$, where[20] $a = (2\pi) 8.1342$ THz and $b = (2\pi) 1.0110$ THz, so the RTPA matrix element can be written

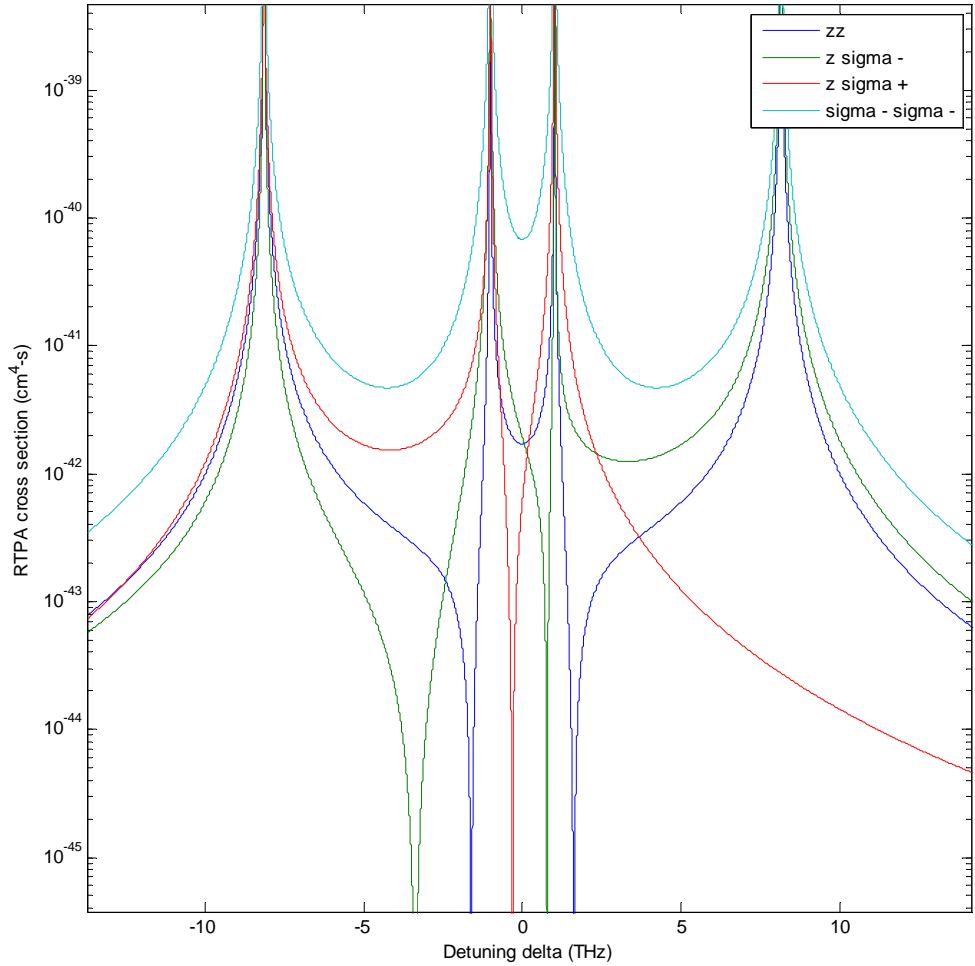

**Figure 2** – Rb random two-photon absorption cross sections for Rb $5S_{1/2} \rightarrow 5D_{3/2}$.

$$M_r = -\left(\frac{D_{21}^{P_{1/2}}}{a-\delta} + \frac{D_{12}^{P_{1/2}}}{a+\delta} + \frac{D_{21}^{P_{3/2}}}{b-\delta} + \frac{D_{12}^{P_{3/2}}}{b+\delta}\right) + M_{other} \qquad (8)$$

$$= -D_{21}^{P_{1/2}}\left(\frac{1}{a-\delta} + \frac{1}{a+\delta} + \frac{\eta}{b-\delta} + \frac{\eta}{b+\delta}\right) + M_{other}$$

where $M_{other}$ represents all other $nP$ intermediate levels ($n \neq 5$) and the continuum. The evaluated cross sections are shown in Figure 2. The double dipole transition matrix elements are given in Table 1. The transition elements are normalized by the empirically derived common factor $R_J = e^2 \left\langle R_{5D_{3/2}} \left| r \right| R_{5P_J} \right\rangle \left\langle R_{5P_J} \left| r \right| R_{5S_{1/2}} \right\rangle \sim 6e^2 a_0^2$ where $R_{nl_j}$ is the reduced radial wave function, derived from the observed Rb transition rates. The $J$-dependence ratio is[21] $\eta \equiv R_{3/2} / R_{1/2} = 1.067(7)$. The cross section is averaged over final states with the density of final states determined by the lifetime $\tau_l \sim 246.4$ ns of the $5D_{3/2}$ state.[22]

The frequency detuning $\delta$ required for BDD destructive interference is obtained by solving $M_r(\delta) = 0$. For $\sigma^- \sigma^-$, $\delta_{\sigma^- \sigma^-} = \pm \sqrt{-ab(\eta a - b)/(a - \eta b)}$, so there is no real solution for $a > 0$, $b > 0$; for $\mathbf{zz}$, $\delta_{\mathbf{zz}} = \pm \sqrt{(5ab^2 + \eta ba^2)/(5a + \eta b)}$; for $\mathbf{z}\sigma^+$, $\delta_{\mathbf{z}\sigma^+} = \frac{1 \pm \sqrt{1 - 20b(\eta-1)(5b+\eta a)/(5\eta a + \eta^2 b)^2}}{10(\eta-1)/(5\eta a + \eta^2 b)}$; for $z\sigma^-$, we solve

$$\delta_{z\sigma^-}^3 + \frac{15a + 7\eta b}{9\eta - 5}\delta_{z\sigma^-}^2 + \frac{5b^2 - 9\eta a^2}{9\eta - 5}\delta_{z\sigma^-} - ab\frac{15b + 7a}{9\eta - 5} = 0.$$

Numeric values for all the BDD detunings are given in Table 1. The deep nulls at the BDD detuning are evident in Figure 2. (The 5P state linewidths $\kappa_P \sim (2\pi)$ 6 MHz have been incorporated in the plot.)

| Pol 1 | Pol 2 | $D_{21}^{1/2}/R$ | $D_{12}^{1/2}/R$ | $D_{21}^{3/2}/(R\eta)$ | $D_{12}^{3/2}/(R\eta)$ | $\delta$ (THz) |
|---|---|---|---|---|---|---|
| $\mathbf{z}$ | $\mathbf{z}$ | $\sqrt{2}/9$ | $\sqrt{2}/9$ | $\sqrt{2}/45$ | $\sqrt{2}/45$ | $\pm 1.645(4)$ |
| $\mathbf{z}$ | $\sigma^+$ | $-1/3\sqrt{3}$ | $0$ | $2/15\sqrt{3}$ | $-1/5\sqrt{3}$ | $-0.3124(7)$ |
| $\mathbf{z}$ | $\sigma^-$ | $1/9$ | $2/9$ | $4/45$ | $-1/45$ | $-3.514(14)$ $+0.7642(6)$ |
| $\sigma^-$ | $\sigma^-$ | $-(2/15)\sqrt{5/6}$ | $-(2/15)\sqrt{5/6}$ | $(2/15)\sqrt{5/6}$ | $(2/15)\sqrt{5/6}$ | No real solution. |

**Table 1** – Double dipole transition matrix elements (normalized by $R = \sim 6e^2 a_0^2$) and destructive interference degeneracy values $\delta$ for candidate BDD polarizations. (Additional far off-resonance solutions that depend strongly on $M_{other}$ are available for $\mathbf{z}\sigma^+$ and $\mathbf{z}\sigma^-$.)

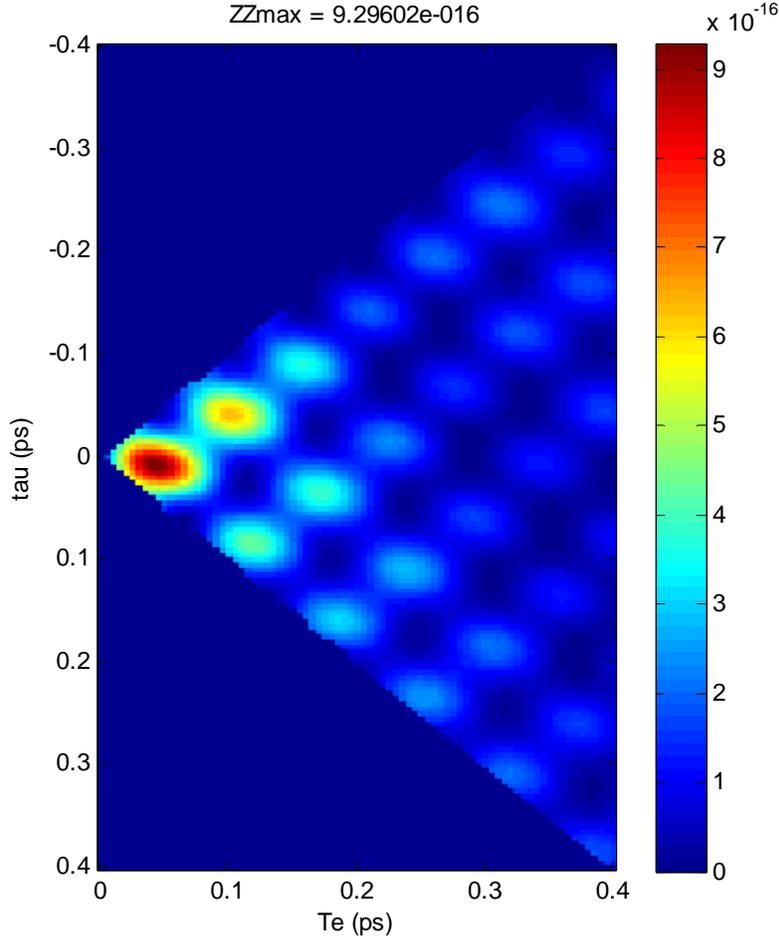

**Figure 3** – ETPA absorption cross section for **zz** polarization with detuning selected to suppress RTPA. The cross section is plotted in cm² as a function of entanglement time $T_e$ and inter-beam delay $\tau$. The maximum value $\sigma_e^{\max} = 10^{-15}$ cm² is achieved at $\tau = 0.0075$ ps and $T_e = 0.043$ ps.

Figure 3 shows the non-zero biphoton absorption as a function of $T_e$ and $\tau$ for **zz**-polarization and BDD detuning $\delta \sim 1.645$THz. To compute the cross section, we must evaluate the entanglement area $A_e$, which depends on the details of the experimental set-up. A typical value can be calculated assuming that the signal and idler beams constitute single modes of their respective optical systems. $A_e$ is then determined by the minimum focal spot diameter achieved by the optical system. The biphotons have wavelength near $\lambda \sim 778$ nm and can be focused onto a focal spot of diameter $D = 2\lambda/(\pi NA)$ for a lens of numerical aperture $NA$, with $NA \sim 1$, $D \sim 500$ nm, and $A_e \sim 2 \times 10^{-9}$ cm². The cross section rapidly oscillates with $T_e$ and $\tau$, with frequency scale set by the intermediate state frequency detuning $a + b \sim 9\text{THz}$. The peak value of the cross section $\sigma_e^{\max} \sim 10^{-15}\ \text{cm}^2$, although this occurs at $T_e = 0.05$ ps.

To summarize, we have shown how entanglement can be used to selectively tag individual quanta, and how interference can be used to differentiate them from otherwise identical particles that are not entangled. We have described an example system based on entangled photons and an Rb vapor cell. BDD can be used in both detection and imaging applications. The advantage of this technique is that it enables nearly complete rejection of background noise.

A number of challenges must be addressed before BDD can be used in realistic systems. Entangled source luminosity must be improved. The two-photon absorption process is weak compared with single photon effects. The narrow-band nature of the random two-photon null is problematic. It may be feasible to address both issues through the use of dressed atomic states similar to those used in slow light and electromagnetically induced transparency.

[1] H. Häffnera, C.F. Roosa, and R. Blatt, "Quantum computing with trapped ions," *Physics Reports* 469, No. 4, 155-203, December 2008.

[2] V. Scarani, H. Bechmann-Pasquinucci, N. Cerf, M. Dušek, and N. Lütkenhaus, "The security of practical quantum key distribution," *Rev. Mod. Phys.* 81, 1301 (2009).

[3] A.N. Boto, P. Kok, D. Abrams, S. Braunstein, C. Williams, and J.P. Dowling, "Quantum Interferometric Optical Lithography: Exploiting Entanglement to Beat the Diffraction Limit," *Phys. Rev. Lett.* 85, 2733 - 2736 (2000).

[4] T.B. Pittman, D.V. Strekalov, D.N. Klyshko, M.H. Rubin, A.V. Sergienko, and Y.H. Shih, "Two-photon geometric optics," *Phys. Rev. A* 53, 2804 - 2815 (1996).

[5] I.F. Santos, J.G. Aguirre-Gómez, S. Pádua, "Comparing quantum imaging with classical second-order incoherent imaging," *Phys. Rev. A* 77, 043832 (2008).

[6] L. Lugiato, A. Gatti, E. Brambilla, "An introduction to quantum imaging," *Quantum Communication and Information Technologies* 113, 85-99 (2003).

[7] Barak Dayan, Avi Pe'er, Asher A. Friesem, and Yaron Silberberg, "Nonlinear Interactions with an Ultrahigh Flux of Broadband Entangled Photons," *Phys. Rev. Lett.* 94, 043602 (2005).

[8] K. Kastella et al., "System and method of detecting entangled photons," US Patent 7,609,382.

[9] S. Lloyd, "Enhanced sensitivity of photodetection via quantum illumination," *Science* 321, No. 5895, 1463-1465, September 12, 2008.

[10] M. Goeppert-Mayer, "Über Elementarakte mit zwei Quantensprüngen," *Ann Phys* 9, 273-95. doi:10.1002/andp.19314010303 (1931).

[11] Albert Gold and H. Barry Bebb, "Theory of Multiphoton Ionization," *Phys. Rev. Lett.* 14, 60-63 (1965).

[12] J. E. Bjorkholm and P. F. Liao, "Resonant Enhancement of Two-Photon Absorption in Sodium Vapor," *Phys. Rev. Lett.* 33, 128-131 (1974).

[13] Hong-Bing Fei, Bradley M. Jost, Sandu Popescu, Bahaa E. A. Saleh, and Malvin C. Teich, "Entanglement-Induced Two-Photon Transparency," *Phys. Rev. Lett.* 78, 1679 - 1682 (1997).

[14] J. Peřina, , Jr., BEA. Saleh, and MC Teich, "Multiphoton absorption cross section and virtual-state spectroscopy for the entangled n-photon state," *Phys. Rev. A* 57, 3972 - 3986 (1998).

[15] B.E.A. Saleh, B.M. Jost, H.B. Fei, M.C. Teich, "Entangled-photon virtual-state spectroscopy," *Phys. Rev. Lett.* 80, No. 16, 3483-3486, April 20, 1998.

[16] G. Mainfray and C. Manus, "Resonance effects in multiphoton ionization of atoms," *Appl. Opt.* 19, 3934-3940 (1980).

[17] Y. H. Shih, A. V. Sergienko, and Morton H. Rubin, "Two-photon entanglement in type-II parametric down-conversion," *Phys. Rev. A* 50, 23-28 (1994).

[18] Morton H. Rubin, David N. Klyshko, Y. H. Shih, and A. V. Sergienko, "Theory of two-photon entanglement in type-II optical parametric down-conversion," *Phys. Rev. A* 50, 5122 - 5133 (1994).

[19] Dong-Ik Lee and Theodore Goodson III, "Entangled Photon Absorption in an Organic Porphyrin Dendrimer," *J. Phys. Chem. B* 110 (51), 25582 – 25585, December 7, 2006.

[20] J. E. Sansonetti, *J. Phys. Chem.* Ref. Data 35, 301 (2006).

[21] S.B.Bayram, M. Havey, M. Rosu, & A. Sieradzan, "$5p^2P_j \rightarrow 5d^2D_{3/2}$ transmission matrix elements in atomic $^{87}$Rb", *Phys. Rev. A* 61, 050502(R) (2000).

[22] D. Sheng, A. Pérez Galván, and L. A. Orozco, "Lifetime measurements of the 5d states of rubidium," *Phys. Rev. A* 78, 062506 (2008).